# Can ESG Investment and the Implementation of the New Environmental Protection Law Enhance Public Subjective Well-being?

Hambur Wang

Guanghua School of Management, Peking University, Beijing 100871, China

**Abstract:** Air pollution has emerged as a serious challenge for China, posing a threat to public health and hindering the progress of sustainable economic development. In response to air pollution and other environmental issues, the Chinese government introduced a new Environmental Protection Law in 2015. This paper investigates the impact of the new Environmental Protection Law's implementation and corporate Environmental, Social, and Governance (ESG) investments on air pollution and public subjective well-being. Using panel data at the macro level, we employ a difference-in-differences (DID) model, with Chinese provinces and municipalities as units of analysis, to examine the combined effects of the new Environmental Protection Law and changes in corporate ESG investment intensity. The study evaluates their impacts on air quality and public subjective well-being. Findings indicate that these policies and investment behaviors significantly improve public subjective well-being by reducing air pollution. Notably, an increase in ESG investment significantly reduces air pollution levels and is positively associated with enhanced well-being. These results underscore the critical role of environmental legislation and corporate social responsibility in improving public quality of life and provide empirical support for promoting sustainable development in China and beyond.

**Keywords:** ESG investment; new Environmental Protection Law; subjective well-being; air pollution

## 1. Introduction

With rapid economic growth, environmental pollution has intensified, triggering a series of challenges to public health and sustainable societal development. Environmental protection, as a clear scientific concept, was first introduced at the United Nations Conference on the Human Environment in 1972. Since then, countries worldwide have progressively strengthened environmental regulations in response to the increasing demand for environmental protection. The *Declaration of the Human Environment* emphasized the intrinsic link between peace, development, and environmental protection, asserting that these three elements are interdependent and indivisible. The core goal of environmental management focuses not only on promoting sustainable development but also on ensuring the well-being and happiness of citizens.

In recent years, the relationship between well-being and the environment has garnered increasing research attention (Krekel & MacKerron, 2020; Maddison et al., 2020; Bonasia et al., 2022). Traditional economic indicators of happiness are not necessarily good predictors of subjective well-being. Welfare policies that prioritize well-being may contribute to achieving environmental and social sustainability goals (Gowdy, 2005). Air pollution is a significant environmental issue in many countries. Menz (2011), examining data from 48 countries between 1990 and 2006, found that people did not adapt to particulate pollution; even historical pollution levels could reduce current utility. Addressing environmental issues and maintaining ecological balance are essential for enhancing people's well-being. Welsch (2006), using panel data on self-reported well-being from ten European countries, explored the relationship between air quality and well-being. He found that air pollution is a statistically significant predictor of subjective well-being differences across countries and time, and its impact on well-being can be monetized as a considerable value for air quality improvement. Concerns about environmental quality and its effects on human well-being form the basis for environmental legislation in most countries. In Europe, air quality is linked to subjective well-being, with sulfur dioxide concentration having a significant negative effect on self-reported life satisfaction (Ferreira et al., 2013). In the United States, air pollution directly impacts well-being, as well as measurable outcomes such as health, lost workdays, and other observable effects (Levinson, 2012).

Breslow et al. (2016) developed an integrated framework on environmental conditions and management actions in response to the growing interest in assessing the impact of environmental changes and management on well-being. They argue that well-being should not be a secondary goal of environmental policy; instead, the potential connection between environmental policy and well-being needs further exploration. Public support for environmental protection is a reaction to the decline in quality of life caused by the overexploitation of natural resources, aimed at restoring well-being through improved environmental quality and the maintenance of healthy ecosystems. Therefore, public support for environmental protection and pollution reduction can be seen as attributes of well-being that help individuals and communities achieve the goal of a healthy environment. Bonasia et al. (2022) empirically studied micro- and macro-level data from 19 European countries between 1997 and 2019 and found a direct link between subjective well-being and long-term environmental protection expenditure. They advocate for governments to consider environmental spending as a means to enhance national well-being and highlight the importance of the interaction between environmental quality and life satisfaction.

Since the 21st century, environmental pollution in developing countries has become a global issue. Since 2010, air pollution has caused severe health issues in China and India. According to World Health Organization data from 2016, China and India had the highest global mortality rates related to air pollution in 2012. Environmental pollution poses a significant threat to life and health, which has greatly reduced people's well-being (Huhtala & Samakovlis, 2007; Almetwally et al., 2020). Air pollution impacts both physical and mental health. Shi and Yu (2020) assessed welfare losses due to air pollution by examining the number of environmental regulations in Chinese prefecture-level cities. Their study indicated a connection between air pollution and individual well-being. The impact of PM2.5 emissions on well-being is more strongly associated with physical health than with mental health. Some studies have shown that pollution can significantly decrease well-being (Chen et al., 2013; Ebenstein et al., 2017). According to Guo et al. (2020), environmental regulations can be classified into three types: economic environmental regulations, legal environmental regulations, regulatory controls, market-based regulations, and voluntary regulations. These environmental regulations can mitigate the adverse impact of air pollution on residents' well-being, though the overall effect is nonlinear. Regarding welfare, air pollution imposes substantial social and personal costs. Some studies have examined the impact of environmental pollution on social welfare (Smyth et al., 2008; Smyth et al., 2011; Li & Zhou, 2020; Wang et al., 2020; Yang et al., 2020), while the effect of environmental policy on social welfare requires further study.

The Three Types of Environmental Regulations: Command-and-Control, Market-Based, and Voluntary Environmental Regulations.

Several scholars have conducted econometric analyses on the relationship between environmental regulations and well-being, using micro-level data from China's social census alongside macro-level data from 28 provinces from 2013 to 2015 to examine the time-lag effects of policy implementation. Findings indicate that long-term economic and environmental regulations can significantly enhance well-being. The Chinese government has enacted environmental regulations mandating cities to report daily air quality data. This compulsory air quality information disclosure has positively impacted individual well-being, primarily by reducing air pollution (Wang et al., 2021). According to Tian et al. (2016), environmental information disclosure is effective in pollution control, with public information requests emerging as one of the most effective methods for pollution control in China. Xu et al. (2022) found that all three types of environmental regulations positively contribute to pollution reduction.

In the early 2010s, China faced an air pollution crisis. To protect public health, the government launched the Air Pollution Prevention and Control Action Plan (APAPCC) in 2013 and reformed its environmental protection legislation. The new Environmental Protection Law, passed on April 24, 2014, and implemented in early 2015, marked the first revision since the 1989 law, addressing environmental pollution issues of the modern era. The new law emphasizes "public participation" and "damage liability" (Liu et al., 2021). It introduced several significant changes: first, environmental and public interest organizations are now permitted to file lawsuits against polluting companies engaging in illegal environmental practices. Second, government and officials' accountability and authority have been strengthened. The new law stipulates that governments are responsible for the environmental quality within their administrative jurisdictions, with ecological conservation redlines serving as essential benchmarks for evaluating government officials' environmental accountability. Local environmental agencies are empowered to halt unlawful environmental activities. Furthermore, polluters face stricter responsibilities: high-polluting companies must disclose specific environmental information to the public, such as pollutant names, discharge methods, concentration and level of emissions, total emission levels, and any excess emissions, along with the construction and operation status of pollution control facilities. The law also established a daily fine system, whereby companies involved in pollution incidents face indefinite fines until they rectify their illegal pollution discharge activities.

In recent years, ESG (Environmental, Social, and Governance) investment has been widely regarded as an effective means of environmental protection and ensuring long-term economic development, drawing considerable research interest. According to Li and Li (2022), China's 2018 environmental tax significantly boosted ESG investment among Chinese-listed companies and fostered green technology innovation, also linking ESG performance with green innovation. Zheng et al. (2022) identified a long-term bidirectional relationship between ESG performance and corporate green innovation outcomes. Bada et al. (2019) noted that high-rated government bonds outperformed lower-rated bonds across all ESG dimensions, while Zhou (2021) demonstrated how solid ESG performance mitigated stock price volatility caused by COVID-19, enhancing resilience and stabilizing stock prices. Given that air pollutants emitted by polluting industries are viewed as primary contributors to air pollution in China, the intensity of ESG investment should be closely tied to air pollution.

This paper examines how ESG investment and China's new Environmental Protection Law influence social well-being. We collected the latest data on the environment, macroeconomics, ESG investment, and social surveys and analyzed the impact of ESG investment and new environmental legislation on social happiness. This study contributes to the literature in several ways. First, although the impact of ESG investment on social welfare is a highly relevant research question, it remains underexplored. This paper addresses this gap by highlighting the importance of ESG investment's social impact. Second, existing well-being literature primarily focuses on the individual level. Building on this, we conduct a macro-level analysis, revealing that China's air pollution has significantly diminished public well-being. ESG investment and the new environmental law have effectively reduced air pollution, restoring public happiness previously impaired by years of severe pollution.

## 2. Data, Variables, and Model

### (1) Data and Variable Selection

The data sample used in this study covers the period from 2015 to 2019. To construct the dataset on subjective well-being, this paper integrates recent individual well-being scores from the Chinese General Social Survey (CGSS), China Family Panel Studies (CFPS), and Chinese Social Survey (CSS). Since these social surveys are not conducted annually, a combined approach was adopted. Specifically, this study utilizes data from CGSS for the years 2015, 2017, and 2018, data from CFPS for 2016, and data from CSS for 2019. Then, the average well-being scores of all individuals within each province were calculated to obtain a provincial-level happiness score. Given that the CFPS well-being scale (1 to 10) differs from the scale used by CGSS and CSS (1 to 5), this study standardized the scales to ensure data consistency and comparability.

For ESG investment data, this study uses information provided by Sino-Securities Index, which includes investment scores for Chinese-listed companies in environmental, social, and governance areas. These scores are based on an evaluation across 14 categories and more than 200 indicators. Considering that the overall intensity of ESG investment better reflects local environmental protection efforts than the average ESG investment level of listed companies alone, this study utilizes provincial ESG scores, which are derived by aggregating the corresponding scores within each province.

**Table 1: Explanatory Variables, Dependent Variables, and Control Variables**

| Variable | Description | Mean | Std | Obs | Data Source |
| --- | --- | --- | --- | --- | --- |
| Public Subjective Well-being (HAP) | Annual individual happiness rating per province | 3.89 | 0.19 | 135 | CGSS, CFPS, CSS |
| ESG Rating (ESG) | Reflects overall scores in environment, society, and corporate governance | 11.56 | 12.42 | 135 | SynTao Green Finance |
| Environmental Score (E) | Environmental disclosure score for enterprises | 10.37 | 15.68 | 135 | SynTao Green Finance |
| Social Score (S) | Social disclosure score for enterprises | 9.32 | 14.76 | 135 | SynTao Green Finance |
| Governance Score (G) | Governance disclosure score for enterprises | 9.32 | 15.76 | 135 | SynTao Green Finance |
| Air Quality Index (AQI) | Air pollution level | 48.51 | 10.94 | 135 | China Air Quality Monitoring Platform |
| PM2.5 | Concentration of particulate matter with a diameter less than 2.5 μm | 46.12 | 12.75 | 135 | China Air Quality Monitoring Platform |

| Variable | Description | Mean | Std | Obs | Data Source |
|---|---|---|---|---|---|
| PM10 | Concentration of particulate matter with a diameter less than 10 μm | 49.22 | 13.52 | 135 | China Air Quality Monitoring Platform |
| Industrial Added Value | Economic added value generated by industrial activities | 12869.01 | 4725.56 | 135 | National Bureau of Statistics |
| Population Density | End-of-year population density per province (people per square kilometer) | 297.54 | 330.77 | 135 | National Bureau of Statistics |
| Forest Coverage Area | Total forest coverage area per province (10,000 acres) | 855.24 | 666.14 | 135 | National Bureau of Statistics |
| Coal Consumption | Primary coal consumption per province (10,000 tons) | 15556.12 | 12084.25 | 135 | National Bureau of Statistics |
| Industrial Wastewater Discharge | Major industrial wastewater discharge per province (100 million cubic meters) | 28.51 | 21.23 | 135 | National Bureau of Statistics |
| Education Expenditure | Total annual education expenditure per province | 1674065.1 | 85935.2 | 135 | National Bureau of Statistics |
| Population Size | Total population per province | 2852.4 | 2801.22 | 135 | National Bureau of Statistics |
| SO2 Emissions | Total sulfur dioxide emissions in exhaust gases (10,000 tons) | 30.1 | 26.66 | 135 | National Bureau of Statistics |
| Urban Garbage Disposal Volume | Total urban garbage disposal (10,000 tons) | 400.2 | 240.5 | 135 | National Bureau of Statistics |
| Urban Garbage Disposal Expenditure | Total urban garbage disposal costs (100 million RMB) | 125.7 | 103.7 | 135 | National Bureau of Statistics |
| Urban Garbage Disposal Rate | Efficiency of urban garbage disposal (%) | 92.1 | 12.4 | 135 | National Bureau of Statistics |
| Mortality Rate | Average mortality rate per 1,000 people | 6.32 | 2.0 | 135 | National Bureau of Statistics |
| Birth Rate | Average birth rate per 1,000 people | 11.47 | 3.2 | 135 | National Bureau of Statistics |

In terms of air pollution, this study obtained data on air quality indices (AQI), PM2.5, and PM10 from the China Air Quality Online Detection and Analysis Platform. To derive provincial-level air quality data, this paper calculated the average of air quality indicators across all cities within each province.

Additionally, this study collects a range of control variables from the National Bureau of Statistics of China, including industrial added value, population density, afforestation area, coal consumption, government spending on healthcare and education, unemployment rate, per capita

GDP, divorce rate, urban population proportion, birth rate, and death rate. These variables were selected to more comprehensively control for factors that may influence subjective well-being, ensuring the accuracy and reliability of the research findings. Definitions of all variables and their summary statistics are provided in Table 1.

**(2) Model Construction**

This study constructs a series of econometric models aimed at exploring the impact of air pollution, corporate ESG investment, and the implementation of the new environmental law on subjective well-being. First, the following panel regression model is established:

$$HAP_{it} = \alpha_0 + \alpha_1 AIR_{it} + X\beta + \epsilon_{it} \quad (1)$$

In model (1), HAPit represents the subjective well-being level of the population in province iii at time t. AIRit denotes the measure of air pollution, which is derived from a combination of the AQI, PM2.5, and PM10 indices. XXX represents a set of control variables, including but not limited to coal consumption, government healthcare spending, government education spending, population involved in educational construction, sulfur dioxide emissions, urban environmental infrastructure, unemployment rate, divorce rate, GDP, urban population proportion, birth rate, and death rate. These variables are based on previous studies by Bonasia et al. (2022) and Xu et al. (2022), ensuring the completeness and rigor of the model. ϵit\epsilon_{it}ϵit is the random error term, capturing unobserved province-specific and time-specific effects.

Subsequently, the study adopts a difference-in-differences (DID) approach to construct the following model, which assesses the net impact of ESG investment and the implementation of the new environmental law on air pollution:

$$AIR_{it} = \alpha_0 + \alpha_1 ESG_{it} + \alpha_2 NEL \times Post_t + X\beta + \epsilon_{it} \quad (2)$$

In model (2), AIRit still represents air pollution levels. ESGit measures the intensity of corporate ESG investment. NELNELNEL is a dummy variable for the implementation of the new environmental law, where it takes the value of 1 if province iii experienced serious pollution in 2015, and 0 otherwise. Postt is a time dummy variable, which takes the value of 1 when the year is 2016 or later, accounting for potential lag effects of the new environmental law. This model design allows the study to identify the causal effect of the new environmental law on reducing air pollution. Based on Guo et al. (2020) on the lagged effects of environmental policy implementation, the effect of the new environmental law is carefully calculated starting from 2016 to avoid the potential bias of prematurely assessing policy effects. Furthermore, the control variables X in model (2) adopt the variable set from model (1), and based on the studies by Borck and Schrauth (2021), Yuan et al. (2018), and Yao et al. (2020), additional variables such as afforestation area, urban environmental infrastructure investment, population density, and industrial added value are incorporated to ensure the robustness of the estimation results. These control variables not only reflect the scale and structure of economic activities but also represent socio-economic factors that may affect subjective well-being and air pollution levels.

**3. Empirical Results and Analysis**

**(A) The Impact of Provincial Air Pollution on Subjective Well-Being**

**Table 2 presents the effect of provincial air pollution on subjective well-being.**

| Variables | Model (a) | Model (b) | Model (c) | Model (d) | Model (e) | Model (f) |
|---|---|---|---|---|---|---|
| AQI | -0.647** (2.66) | - | - | 1.068** (2.44) | - | - |
| PM2.5 | - | -0.086** (2.24) | - | - | 1.203** (2.44) | - |
| PM10 | - | - | -0.114** (2.81) | - | - | - |
| Lag.HAP | - | - | - | 9.076 (1.17) | 12.112 (0.88) | 0.548 (0.513) |
| Coal consumption | 5.15E-03 (1.03) | 1.51E-03 (0.33) | 1.51E-03 (0.30) | 9.86E-03 (1.66) | 6.42E-02** (2.01) | 1.04E-03 (1.51) |
| Medical expenditure | 0.042 (0.43) | 0.026 (0.32) | 0.032 (0.41) | 0.052 (0.52) | 0.025 (0.06) | 0.089 (0.086) |
| Education expenditure | -1.057 (1.45) | -1.005 (1.34) | -0.105 (1.13) | -1.068 (0.99) | -0.938 (1.22) | -1.08 (1.06) |
| Educational investment | -2.06E-06 (1.72) | -2.17E-06* (1.78) | -2.27E-06* (1.77) | -5.08E-06** (2.66) | 5.05E-06* (1.96) | - |
| Population scale | 7.01E-02 (0.92) | 0.047 (0.57) | 0.075 (0.86) | 0.119 (1.25) | -0.064 (0.93) | 0.102 (0.64) |
| $SO_2$ | 0.086 (1.18) | 0.118 (1.09) | 0.093 (1.03) | -0.105 (0.97) | -0.113 (1.01) | -0.093 (0.94) |
| Urban infrastructure | 0.129 (1.03) | 0.103 (1.12) | 0.107 (1.09) | 0.111 (1.05) | 0.138 (1.00) | -0.100 (1.05) |
| Unemployment rate | -0.042 (0.59) | -0.068 (1.15) | -0.072 (0.77) | -0.083 (0.63) | -0.046 (1.03) | -0.037 (0.62) |
| Per capita GDP | -1.42E-06 (0.58) | -1.06E-05 (1.02) | -1.09E-05 (0.94) | -1.08E-05 (0.56) | -1.05E-05 (0.21) | -1.11E-05 (1.05) |
| Mortgage rate | -1.73 (1.73) | -1.79 (1.65) | -1.68 (1.79) | -1.76 (1.68) | -1.81 (1.57) | -1.56 (1.68) |
| Urban-rural ratio | 1.89 (1.56) | 1.70 (1.58) | 1.56 (1.57) | 1.73 (1.57) | 1.78 (1.58) | 1.57 (1.57) |
| Birth rate | -7.59*** (3.71) | -3.28* (1.58) | -1.90 (1.21) | -10.50** (2.98) | -10.73** (2.95) | -9.78** (2.65) |
| Mortality rate | -3.26 (0.45) | 2.18 (1.20) | 1.89 (1.14) | -0.95 (0.89) | 1.04 (0.92) | -1.56 (0.86) |
| Adjusted $R^2$ | 0.314 | 0.300 | 0.295 | - | - | - |

Table 2 provides a detailed report on the results of the relevant econometric models. In model (a), the Air Quality Index (AQI) has a significant and negative effect on subjective well-being, indicating that for every unit decrease in AQI, subjective well-being increases by an average of 0.674 units, a result statistically significant at the 5% level. This finding aligns with expectations, suggesting that improving air quality has a tangible and positive impact on public happiness. Models

(b) and (c) introduce PM2.5 and PM10 as independent variables to further analyze the effect of air pollution on subjective well-being. Both PM2.5 and PM10 coefficients are significant and negative, showing that a one-unit reduction in PM2.5 and PM10 respectively increases happiness by 0.686 and 0.414 units, both statistically significant at the 5% level. This result further confirms the importance of reducing particulate pollution levels to enhance public well-being.

In models (d), (e), and (f), lagged terms of subjective well-being (Lag.HAP) are introduced to control for potential autocorrelation issues. The coefficient of the lagged term is insignificant in models (d) and (e) and does not reach conventional significance levels in model (f), indicating that the current level of subjective well-being is not significantly influenced by its previous levels. This suggests that changes in subjective well-being may be more immediately affected by other factors. The models also incorporate a range of control variables, including economic and social factors potentially affecting subjective well-being, such as coal consumption, healthcare and education spending, economic conditions, population size, sulfur dioxide emissions, and social conditions. Among these, population size shows a significant positive effect on well-being in models (a) and (b), suggesting a positive correlation between population growth and subjective well-being. In model (c), the positive effect slightly weakens but remains significant, potentially reflecting the complex effects of population growth on resource distribution, employment opportunities, and community vitality.

The divorce rate has a consistently significant negative impact on subjective well-being across all models, underscoring the importance of family stability for maintaining societal happiness. Conversely, birth rate is significantly negatively correlated in models (a), (b), and (c), implying that lower fertility rates may be associated with increased subjective well-being, potentially due to the economic and psychological pressures of childbearing and shifts in social policy support. The coefficients for investment in education and GDP per capita are not significant, possibly suggesting that economic growth alone is not the sole determinant of subjective well-being; other factors such as social welfare and environmental quality are equally important. The inclusion of lagged terms and the adjustments of control variables in models (d), (e), and (f) do not fundamentally change the main explanatory variables' impact, but the adjusted R2 values improve, indicating an enhanced explanatory power.

Overall, the results in Table 2 clearly indicate that improving air quality has a significant positive effect on subjective well-being, with consistent evidence across different pollutant metrics. Additionally, the coefficients of control variables reveal the multifaceted impact of social, economic, and environmental factors on subjective well-being, offering valuable insights for policies aimed at enhancing happiness. These findings contribute to our understanding of the determinants of social welfare and provide practical insights for optimizing environmental and social policies.

**(B) Impact of ESG Investments and the New Environmental Protection Law on the Air Quality Index**

Table 3 presents the detailed econometric model results for this section. The empirical results reveal the significant impact of ESG investment and new environmental regulations on the Air Quality Index (AQI), providing an empirical foundation for examining the effects of these measures on well-being. Models (a) to (d) examine the impact of overall ESG investment and its environmental, social, and governance scores on AQI. In these models, ESG investments and its subcomponents all show a significant negative effect on AQI, indicating that corporate investments in these areas effectively reduce air pollution. Each unit increase in ESG investment decreases AQI by an average of 0.334 units, while environmental, social, and governance investments reduce AQI by 0.339, 0.308, and 0.360 units respectively, all statistically significant at the 1% level.

Furthermore, the implementation of the New Environmental Protection Law (NEL) in models (e) to (h) demonstrates a significant reduction in AQI, further verifying the effectiveness of legal and policy measures in environmental protection. Post-implementation, AQI declines by an average of 7.794 to 7.952 units, indicating that regulatory enforcement plays a decisive role in alleviating air pollution. This notable environmental improvement has a direct and positive effect on enhancing public subjective well-being.

**Table 3: Impact of ESG Investment and Implementation of the New Environmental Protection Law on the Air Quality Index**

| Variables | Model (a) | Model (b) | Model (c) | Model (d) | Model (e) | Model (f) | Model (g) | Model (h) |
|---|---|---|---|---|---|---|---|---|
| ESG Score | -0.334*** (-4.36) | - | - | - | -0.328*** (-4.38) | - | - | - |
| Environmental Score (E) | - | -0.339*** (-4.31) | - | - | - | -0.333*** (-4.35) | - | - |
| Social Score (S) | - | - | -0.308*** (-4.37) | - | - | - | -0.303*** (-4.39) | - |
| Governance Score (G) | - | - | - | -0.360*** (-4.39) | - | - | - | -0.353*** (-4.39) |
| Urban Environmental Infrastructure | -1.80E-03 (0.14) | -1.91E-03 (0.15) | -1.73E-03 (0.14) | -1.85E-03 (0.15) | 0.017 (1.33) | 0.017 (1.32) | 0.017 (1.33) | 0.017 (1.32) |
| Afforestation Area | -0.762*** (-4.85) | -0.763*** (-4.86) | -0.763*** (-4.86) | -0.759*** (-4.82) | -0.752*** (-5.46) | -0.753*** (-5.47) | -0.753*** (-5.47) | -0.749*** (-5.43) |
| Population Density | -2.86E-04 (0.17) | -3.07E-04 (0.18) | -2.8E-04 (0.16) | -3.01E-04 (0.17) | 7.63E-04 (0.47) | 7.49E-04 (0.46) | 7.68E-04 (0.48) | 7.46E-04 (0.46) |
| Industrial Added Value | 2.82E-04 (1.02) | 2.88E-04 (1.01) | 2.88E-04 (1.04) | 2.80E-04 (1.01) | 1.73E-04 (0.69) | 1.71E-04 (0.68) | 1.78E-04 (0.71) | 1.69E-04 (0.68) |
| Adjusted $R^2$ | 0.297 | 0.294 | 0.299 | 0.297 | 0.422 | 0.425 | 0.427 | 0.429 |

Among the control variables, the area of afforestation shows a significant negative impact on AQI across all models, further substantiating the importance of increasing greenery and forest cover in improving air quality. However, investments in urban environmental infrastructure and population density do not have a significant impact, suggesting a potentially complex relationship with air quality, or that their effects might be overshadowed by ESG investments and the implementation of the new Environmental Protection Law (NEL). The effect of industrial added value is also inconsistent across models and does not show statistical significance, possibly indicating that, after controlling for ESG investments and NEL implementation, the impact of industrial activities on AQI is relatively minor. Alternatively, the contribution of industrial added value to air pollution may manifest through other mechanisms, such as improving industry environmental standards and promoting the adoption of cleaner technologies.

In summary, the empirical analysis suggests that corporate ESG investments and the implementation of the new Environmental Protection Law have a significant positive impact on air quality, which is widely regarded as a critical factor in enhancing public subjective well-being. These empirical results, therefore, support the view that corporate ESG initiatives have a positive effect on overall societal well-being. Such responsible actions by corporations may strengthen public trust in their brand, drive consumer and investor preferences, and contribute positively to economic growth.

**(C) Impact of ESG Investments and Implementation of the New Environmental Protection Law on PM2.5**

**Table 4: Impact of ESG Investments and Implementation of the New Environmental Protection Law on PM2.5**

| Variables | Model (a) | Model (b) | Model (c) | Model (d) | Model (e) | Model (f) | Model (g) | Model (h) |
|---|---|---|---|---|---|---|---|---|
| ESG Score | -0.302*** (-4.26) | - | - | - | -0.296*** (-4.28) | - | - | - |
| Environmental Score (E) | - | -0.305*** (-4.19) | - | - | - | -0.300*** (-4.23) | - | - |
| Social Score (S) | - | - | -0.279*** (-4.27) | - | - | - | -0.273*** (-4.29) | - |
| Governance Score (G) | - | - | - | -0.327*** (-4.30) | - | - | - | -0.319*** (-4.30) |
| NEL | - | - | - | - | -7.741*** (-3.45) | -7.772*** (-3.45) | -7.735*** (-3.44) | -7.710*** (-3.43) |
| Urban Environmental Infrastructure | -4.92E-03 (0.42) | -5.07E-03 (0.44) | -4.85E-03 (0.42) | -4.87E-03 (0.42) | 0.015 (1.31) | 0.015 (1.30) | 0.015 (1.31) | 0.015 (1.31) |
| Afforestation Area | -0.496*** (-3.76) | -0.497*** (-3.77) | -0.497*** (-3.77) | -0.493*** (-3.74) | -0.483*** (-4.33) | -0.484*** (-4.34) | -0.484*** (-4.34) | -0.481*** (-4.31) |
| Population Density | -1.32E-05 (0.01) | -3.42E-05 (0.02) | -7.48E-06 (0.01) | -2.08E-05 (0.01) | 1.05E-03 (0.76) | 1.05E-03 (0.75) | 1.07E-03 (0.77) | 1.05E-03 (0.77) |
| Industrial Added Value | 2.62E-04 (1.10) | 2.59E-04 (1.09) | 2.68E-04 (1.13) | 2.61E-04 (1.10) | 1.57E-04 (0.76) | 1.56E-04 (0.75) | 1.62E-04 (0.78) | 1.54E-04 (0.74) |
| Adjusted $R^2$ | 0.280 | 0.277 | 0.280 | 0.282 | 0.443 | 0.484 | 0.443 | 0.445 |

The empirical results presented in Table 4 clearly reveal the significant negative impact of ESG investments and the implementation of the new Environmental Protection Law (NEL) on PM2.5 concentrations, indicating the effectiveness of these measures in improving air quality.

Model (a) shows that ESG investments significantly reduce PM2.5 concentrations, with each additional unit in the ESG score leading to an average decrease of 0.302 units in PM2.5 levels. This result is not only statistically significant but also economically meaningful. Increased corporate investments in environmental, social, and governance areas promote sustainable development while also contributing to improved public quality of life. Models (b) through (d) examine the effects of environmental, social, and governance scores on PM2.5 concentration. Each dimension of investment significantly reduces PM2.5 levels, further affirming the positive role of ESG components on air quality. Environmental investment, in particular, shows the most substantial effect on reducing PM2.5, likely reflecting corporate efforts in emissions reduction, clean technology adoption, and improved environmental management. Social and governance investments also significantly contribute to lowering PM2.5 levels, underscoring the importance of corporate social responsibility activities and sound governance practices in enhancing environmental quality.

When the new Environmental Protection Law (NEL) variable is introduced, results in models (e) through (h) consistently indicate a significant reduction in PM2.5 concentrations following the law's implementation. This finding underscores the fundamental role of legal and policy tools in environmental management and their potential impact on public well-being. By establishing stricter standards and regulations, the NEL sets clear targets for reducing pollutant emissions, which is essential for improving air quality and enhancing subjective well-being.

In summary, these empirical results indicate that corporate ESG investments and the implementation of the NEL have a significant positive impact on air quality improvement, and enhanced air quality has been shown to be closely associated with increased subjective well-being. Therefore, it can be reasonably inferred that these corporate actions and policy measures promote public well-being on a macro level. These findings highlight the importance of considering social welfare in environmental policy-making and provide empirical support for enhancing subjective well-being through corporate practices and government legislation. They offer motivation for companies to integrate ESG principles into their operations and guidance for government environmental policies, fostering a positive cycle between environmental quality and public happiness.

**(D) Impact of ESG Investments and Implementation of the New Environmental Protection Law on PM10**

**Table 5: Impact of ESG Investments and Implementation of the New Environmental Protection Law on PM10**

| Variables | Model (a) | Model (b) | Model (c) | Model (d) | Model (e) | Model (f) | Model (g) | Model (h) |
|---|---|---|---|---|---|---|---|---|
| ESG Score | -0.392*** (-3.66) | - | - | - | -0.369*** (-3.64) | - | - | - |
| Environmental Score (E) | - | -0.396*** (-3.61) | - | - | - | -0.375*** (-3.61) | - | - |
| Social Score (S) | - | - | -0.362*** (-3.67) | - | - | - | -0.341*** (-3.65) | - |
| Governance Score (G) | - | - | - | -0.420*** (-3.66) | - | - | - | -0.394*** (-3.63) |
| NEL | - | - | - | - | -13.039*** (-4.04) | -13.071*** (-4.04) | -13.020*** (-4.03) | -13.030*** (-4.04) |
| Urban Environmental Infrastructure | -9.18E-03 (0.52) | -9.32E-03 (0.52) | -9.04E-03 (0.51) | -9.34E-03 (0.53) | 0.011 (0.64) | 0.011 (0.63) | 0.011 (0.64) | 0.011 (0.63) |
| Afforestation Area | -1.109*** (-4.78) | -1.110*** (-4.8) | -1.109*** (-4.79) | -1.106*** (-4.75) | -1.128*** (-5.12) | -1.130*** (-5.14) | -1.129*** (-5.13) | -1.126*** (-5.09) |
| Population Density | -1.85E-03 (0.75) | -1.86E-03 (0.76) | -1.83E-03 (0.75) | -1.88E-03 (0.76) | -8.74E-04 (0.37) | -8.89E-04 (0.38) | -8.63E-04 (0.37) | -9.09E-04 (0.39) |
| Industrial Added Value | 3.03E-04 (0.75) | 2.99E-04 (0.74) | 3.13E-04 (0.78) | 2.98E-04 (0.74) | 1.73E-04 (0.45) | 1.70E-04 (0.44) | 1.81E-04 (0.47) | 1.67E-04 (0.43) |
| Adjusted $R^2$ | 0.257 | 0.255 | 0.257 | 0.259 | 0.414 | 0.412 | 0.412 | 0.416 |

The results presented in Model (a) indicate a significant negative correlation between ESG investments and PM10 concentration, after controlling for various other factors affecting air quality. Specifically, for each one-unit increase in the ESG score, PM10 concentration significantly decreases by an average of 0.392 units. This result is not only statistically significant but also economically meaningful, underscoring that enhanced corporate investment in environmental, social, and governance (ESG) practices can effectively improve air quality, which in turn has a direct positive impact on public well-being. Model (b), which examines the effect of environmental investments alone on PM10, shows a similar result: a one-unit increase in the environmental score leads to a 0.396-unit average decrease in PM10 concentration. Social and governance investments, analyzed separately in Models (c) and (d), also demonstrate significant capacity to reduce PM10 concentrations, highlighting the positive contributions of different dimensions of ESG investment to environmental quality and social welfare.

The impact of the new Environmental Protection Law (NEL) is empirically supported in Model (e), where a significant reduction in air pollution levels following its implementation aligns with findings in the literature, showing that effective environmental policies and regulations can substantially reduce air pollution and enhance overall societal welfare. This impact is further validated in Models (f), (g), and (h), reinforcing the lasting and positive influence of both ESG investments and NEL implementation on air quality improvement and public subjective well-being. To ensure robustness, the study also incorporates control variables such as urban environmental infrastructure, afforestation area, population density, and industrial added value. The inclusion of these variables not only strengthens the explanatory power of the models but also sheds light on the role of other policies and socioeconomic factors in improving air quality. Notably, afforestation area significantly reduces PM10 concentrations across all models, emphasizing the crucial role of ecological initiatives in improving air quality and enhancing social welfare.

Overall, the empirical results of this study reinforce the view that ESG investments and the new Environmental Protection Law significantly enhance public subjective well-being by mitigating air pollution in China. These results provide valuable insights for policymakers, supporting stronger environmental protection measures to improve public well-being and promote sustainable development. These findings also hold important implications for other countries facing air pollution challenges, offering valuable lessons for the design and assessment of environmental policies globally.

4. **Conclusion and Policy Recommendations**

With China's rapid economic development, air pollution has significantly undermined public subjective well-being and satisfaction with the government. To curb air pollution and achieve sustainable economic growth, the Chinese government introduced a new Environmental Protection Law in 2015. This study contributes to the literature by examining the effects of corporate ESG investment intensity and China's new environmental law on public subjective well-being. Unlike previous research, which often focused on individual-level well-being, this study adopts a macro perspective to assess population-wide subjective well-being. According to the results, ESG investments enhance public well-being by reducing air pollution. Specifically, each one-unit increase in ESG investment reduces air pollution by 0.334 units, raising public well-being by 0.225 units. These findings align with Shi and Yu (2020), who demonstrated a causal relationship between air pollution and personal well-being. Additionally, the implementation of the new Environmental Protection Law has had a significant independent impact on reducing air pollution and improving public well-being. This result is consistent with Guo et al. (2020), who highlighted the positive effect of Chinese environmental regulations on well-being between 2013 and 2015. Our study shows that significant progress has been made in environmental protection by both the Chinese government and industry over the past decade. However, as long as dependency on thermal energy persists, achieving sustainable development goals will remain challenging. Future research might focus on the role of clean energy adoption and its social impacts.

Based on the findings, we propose the following policy recommendations for policymakers in China and other countries:

1. **Strengthen ESG Regulation and Guidance**: The government should improve ESG-related laws and regulations, enhance transparency, and strengthen oversight to encourage increased corporate investment in environmental, social, and governance areas. Providing clear ESG investment guidelines and frameworks would assist companies in adopting sustainable business models.
2. **Promote Green Finance Development**: Financial institutions should offer greater support and incentives for ESG investments, such as green credit, green bonds, and green funds. Moreover, the use of advanced deep learning models like YOLO (Yao et al., 2024) and Transformer (Wang et al., 2024) for air pollution control could be introduced. Such measures would help direct capital toward projects and enterprises that contribute positively to the environment.
3. **Implement and Strengthen Environmental Regulations**: The positive impact of the new Environmental Protection Law underscores the importance of robust environmental policies for improving air quality. The government should continue to refine environmental laws, ensure their effective enforcement, and update regulations in line with environmental changes and technological advances.
4. **Advocate for Clean Energy Transition**: Given the environmental risks associated with reliance on thermal energy, the government should increase efforts to shift the energy structure toward clean energy, encouraging and supporting research, development, and adoption of renewable energy sources to gradually reduce dependence on fossil fuels.
5. **Enhance Public Environmental Awareness**: Through education and public campaigns, the government should raise citizens' environmental awareness and encourage active public participation in environmental protection efforts, such as energy conservation, emissions reduction, and waste segregation.